\documentclass[10pt]{amsart}

\usepackage{amscd,amsthm,mathrsfs,eucal,dsfont,verbatim}
\usepackage[english]{babel}
\usepackage[left=3cm,right=3cm,top=2.5cm,bottom=3cm]{geometry}

\usepackage{amsmath,amssymb,amsthm,bbm,color,graphics,version}
\usepackage{mathrsfs}
\usepackage{mathtools}
\usepackage{amsfonts, xfrac}
\usepackage{enumitem, listings, color}
\usepackage{listings}
\usepackage{geometry,array}

\usepackage{lscape}
\usepackage{color}
\usepackage{setspace}

\usepackage{hyperref}

\theoremstyle{plain}
\newtheorem{theorem}{Theorem}
\newtheorem{lemma}[theorem]{Lemma}
\newtheorem{corollary}[theorem]{Corollary}

\theoremstyle{definition}
\newtheorem{definition}[theorem]{Definition}
\newtheorem{example}[theorem]{Example}

\newcommand{\po}{\ensuremath{\operatorname{d}\!}}
\newcommand*{\QEDA}{\hfill\ensuremath{\blacksquare}}

\definecolor{dkgreen}{rgb}{0,0.6,0}
\definecolor{gray}{rgb}{0.5,0.5,0.5}
\definecolor{mauve}{rgb}{0.58,0,0.82}

\usepackage{color}

\usepackage{soul}

\title{Modeling social media contagion using Hawkes processes}

\author[Z. Palmowski]{Zbigniew Palmowski}
\address{ Faculty of Pure and Applied Mathematics,
Wroc\l aw University of Science and Technology,
Wyb. Wyspia\'nskiego 27, 50-370 Wroc\l aw, Poland}
\email{zbigniew.palmowski@pwr.edu.pl}
\author[D. Puchalska]{Daria Puchalska}
\address{ Faculty of Pure and Applied Mathematics,
Wroc\l aw University of Science and Technology,
Wyb. Wyspia\'nskiego 27, 50-370 Wroc\l aw, Poland}
\email{puchalska.daria@wp.pl}

\thanks{This work is partially supported by Polish National Science Centre Grant No. 2016/23/B/HS4/00566, 2017-2020}

\date{\today}

\begin{document}

\begin{abstract}
The contagion dynamics can emerge in social networks when repeated activation is allowed.
An interesting example of this phenomenon is retweet cascades where
users allow to re-share content posted by other people with public accounts.
To model this type of behaviour we use a Hawkes self-exciting process.
To do it properly though one needs to calibrate model under consideration.
The main goal of this paper is to construct moments method of estimation
of this model. The key step is based on identifying of a generator of
a Hawkes process. We perform numerical analysis on real data as well.

\vspace{3mm}


\end{abstract}

\maketitle

\pagestyle{myheadings} \markboth{\sc D.\ Puchalska --- Z.\ Palmowski
} {\sc Modeling social media contagion using Hawkes processes}

\vspace{1.8cm}

\tableofcontents

\newpage

\section{Introduction}\label{sec:iar}
Nearly every day one can hear about situations in which a single incident causes a series of similar events in a short time interval. Multiple aftershocks or increased seismic activity not only in the place of the earthquake occurrence, but also in surrounding locations, are one of the most common examples. Another one can be a crash on the US stock market influencing the rest of the world’s financial markets. This kind of events are often called a cascade effect. This type of phenomenon can be observed in social media as well
when repeated activation is allowed. In this paper we focus on retweet cascades where
users allow to re-share content posted by other people with public accounts.

The first main goal of this paper is to model this type of behaviour by a Hawkes process.
This process was proposed by Hawkes \cite{hawkes_original}
to describe the dynamics changing in time in which every event causes an increase in the probability of next events occurring.
Although the term ''Hawkes process'' has been used in scientific literature since the early 1970s, the model gained popularity only in the last decade. Its popularity is gained by its clustering property.

The second goal of this paper is to construct calibration method of the Hawkes process based on identifying its
infinitesimal generator. Then we show how to apply it in modelling retweet cascades.
This work combines elements of the probability theory, statistics, the theory of stochastic processes and computer simulations.

Modelling using Hawkes have been widely used in many areas. The moments method in this context was described
in \cite{CHY}. Similar considerations have been performed in the context
queues driven by Hawkes processes in \cite{DP1}, where generator method has been also applied.
This method can be also applied for other Markov process whose
moments are easy to compute by exploiting the structure of a new sequence of nested matrices called
Matryoshkhan matrices; see \cite{DP2} for details.
One can work with generalization of the Hawkes process as well called the ephemerally
self-exciting process; see \cite{DP3}.
The connection of the Hawkes process with social media has not been yet understood well though, in our opinion.
The first attempt can be found in \cite{hawkes_in_social_media}. The
contagion dynamics in social media have been observed and confirmed in many papers and in many contexts, see e.g.
\cite{1,2,3,4,5,6,7}
and references therein.

The paper is organized as follows.
In Section \ref{chap1} we present main theoretical facts and we give there our first main result
concerning generator of the Hawkes process.
In next section we present estimation method of the Hawkes processes parameters.
In Section \ref{sectwitter} we show how to model retweet cascades using previous analysis.
Finally, Section \ref{secconcl} concludes this paper.
All proofs are shifted to Appendix to gain the readability of main message of this work.

\section{Hawkes process and its generator}\label{chap1}
The (Markov) Hawkes process $N_t$ was first time defined by Alan G. Hawkes in 1971 in his seminal work~\cite{hawkes_original} as a~counting process which intensity
\begin{equation*}
    \lambda_t=\lim_{h\to0}\frac{\mathbb{E}\big[N_{t+h}-N_t\,|\,\mathcal{F}_t\big]}{h}\,,
\end{equation*}
at the moment $t$ depends on its natural history $\mathcal{F}_t$
till that moment, that is, it solves the following stochastic differential equation
\begin{equation}\label{eq:d_lambda}
\po\lambda_t=\beta(\lambda_{\infty}-\lambda_t)\po t + \alpha\po N_t\,,
\end{equation}
where $\beta>\alpha>0$ and $\lambda_{\infty}>0$ is a fixed constant called the \textit{base intensity}.
Note that
\begin{equation}\label{eq:lambda_wyprowadzony_wzor}
    \lambda_t=\lambda_{\infty}+(\lambda_0-\lambda_{\infty})\,e^{-\beta t}+\alpha\int_0^t e^{-\beta(t-s)}\po N_s
    =
\lambda_\infty+(\lambda_0-\lambda_\infty)\,e^{-\beta t} + \sum_{0\leq T_k<t}\alpha\,e^{\beta(t-T_k)},
\end{equation}
where $\{T_k\}_{k=1,2,\ldots}$ are consecutive moments of jumps of $N_t$. Moreover,
we have that $N_0=0$, $\mathbb{P}(N_{t+h}-N_t=0\,|\,\mathcal{F}_t)=1-\lambda_t\,h + o(h)$,
$\mathbb{P}(N_{t+h}-N_t=1\,|\,\mathcal{F}_t)=\lambda_t\,h + o(h)$
and $\mathbb{P}(N_{t+h}-N_t>1\,|\,\mathcal{F}_t)=o(h),$ where $o(h)$ denotes quantity tending to zero as $h$ tends to zero.
Therefore, each jump of the process $N_t$ brings an increase in its intensity function by value $\alpha$ which increases
in turn the likelihood of another jump. This is why this process is called a \textit{self-exciting process}. Then the intensity function decays exponentially to the base level $\lambda_\infty$ at a rate given by the factor $\beta$. The condition $\beta>\alpha$ is necessary to avoid the so-called process explosion.

\begin{figure}[ht]
    \centering
    \includegraphics[width=15cm]{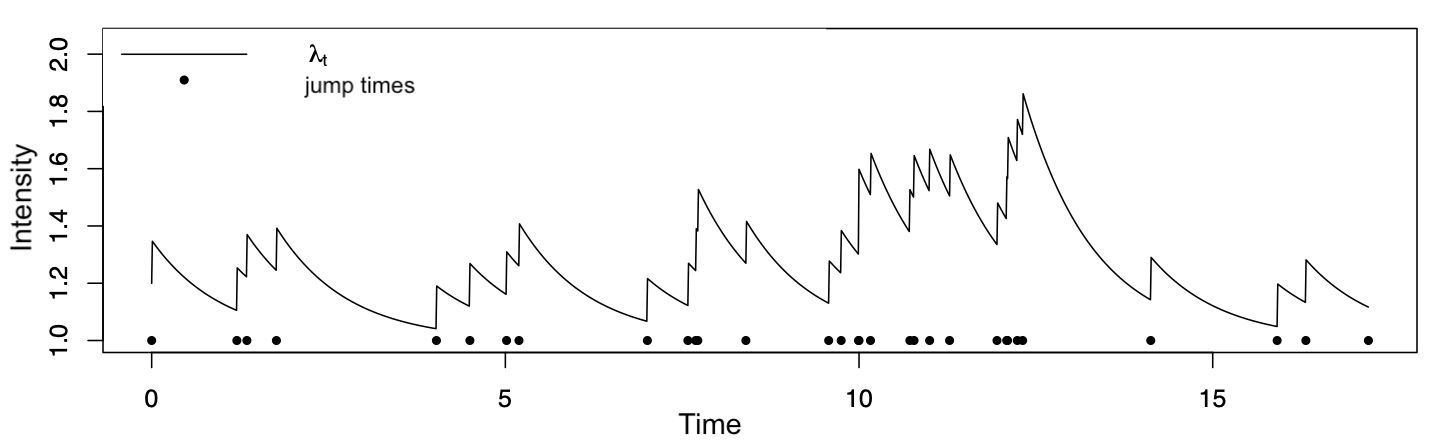}
    \caption{Example of the intensity function trajectory for $\lambda_0=1.2,\,\lambda_\infty=1,\,\alpha=0.15,\,\beta=1.$}
    \label{fig:lambda_t}
\end{figure}

The Hawkes process simulation algorithm was proposed by Dassios and Zhao \cite{dassios_simulation}. The algoritm presented in their work can be implemented in~the~\textit{R} programming language (see Algoritm \ref{kod} in Appndix) and one of~its exemplary outcomes is shown in Figure~\ref{fig:lambda_t}.

The clustering effect was not directly visible in the definitions and formulas cited so far, although it is a distinctive feature of the Hawkes process. For this reason, we recall an alternative equivalent definition of this process
which is based on the Poisson process construction with clustering, where clusters are formed according to a recursive branching structure.
More formally,
Hawkes process with the exponentially decaying intensity is a Poisson cluster process.
The cluster centers of~$C$ are the particular points called immigrants, the rest of the points are called offspring.
The Hawkes process can be constructed as follows:
\begin{enumerate}[label=(\alph*)]
    \item the immigrants are coming at moments $\widetilde{T}_m$ according to inhomogeneous Poisson process with rate $\lambda_\infty+(\lambda_0-\lambda_\infty)\,e^{-\beta t}$, $t\geq0$;
    \item each immigrant who arrived at $\widetilde{T}_m$ generates one cluster $C_m$, and all these clusters are independent of each other;
    \item each cluster $C_m$ is a random set formed by points of generations of order $n=0,1,\ldots$ with the following branching structure:
    \begin{itemize}
        \item the immigrant is said to be of generation 0;
        \item recursively, given generations $0,1,\ldots,n$ in $C_m$, each $\widetilde{T}_j\in C_m$ of generation $n$ generates a Poisson process of offspring of generation $n+1$ on $(\widetilde{T}_j,\infty)$ with intensity $\alpha e^{-\beta(t-\widetilde{T}_j)},\,t>\widetilde{T}_j$, independent of generations $0,1,\ldots,n$;
    \end{itemize}
    \item $C$ consists of the union of all clusters, i.e. $C=\bigcup_{m=1,2,\ldots}C_m$ (see \cite{dassios_simulation}).
\end{enumerate}

\begin{figure}[h]
    \centering
    \includegraphics[width=12cm]{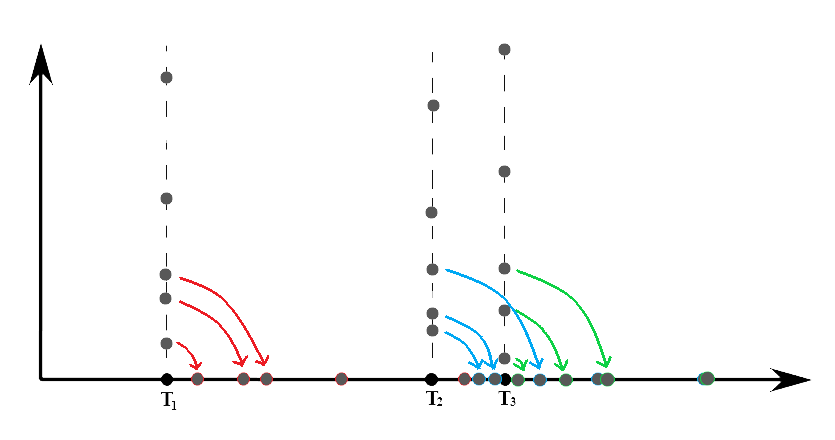}
    \caption{Simplified diagram of the Hawkes process generation.}
    \label{fig:schemat}
\end{figure}

This means that first the generation of the so-called "immigrants" is generated. These are jump moments of the non-stationary Poisson process with the intensity function equal to the deterministic part of the expression in \eqref{eq:lambda_wyprowadzony_wzor}. For each of them, the Poisson process is generated again with the intensity specified in the definition and with a shift in time, i.e. for $t>T_j$. These points are projected onto the Hawkes process timeline and treated as its jump moments. This algorithm is repeated for each of the generations. The simulation scheme for the first two generations is presented in Figure~\ref{fig:schemat}.

Successive generation points appear with exponentially decaying intensity, so that most of them, when projected onto the same timeline, gather near the parent. This is an obvious reason for the creation of the clusters we observe.

The first aim of this paper is to
construct a new method of calibration of the Hawkes process $N_t$.
To do so we will find its (infinitesimal) generator.

From \cite{hawkes_is_markovian} we know that the two-dimensional process $X_t=(\lambda_t,\,N_t)$ is a Markov process on $D=\mathbb{R}_+\times\mathbb{N}\cup\{0\}$.
For a sufficiently regular function $k:D\rightarrow\mathbb{R}$ we define the generator of $X_t$ as the following limit
\begin{equation*}
    \mathcal{A}\,k(x) = \lim_{h\to0}\frac{\mathbb{E}_t^x [k(X_{t+h})]-k(x)}{h},
\end{equation*}
with $\mathbb{E}_t^x[\quad\cdot\quad]=\mathbb{E}^x [\quad\cdot\quad|\mathcal{F}_t\quad {\rm and} \quad X_t=x]$.
We will call the family of functions, for which the limit in the definition of the generator makes sense, the domain of the generator $\mathcal{A}$ and denote by $\mathcal{D}(\mathcal{A})$.
From Dynkin theorem we know that $
    M_t = k(X_t) - k(X_0)-\int_0^t\mathcal{A}k(X_s)\po s,\,t\geq0
$
is a martingale with respect to the filtration $\mathcal{F}_t$. Thus from Optional Stopping Theorem
we can conclude that
\begin{equation}\label{keyidenity}
    \mathbb{E}\big[k(X_t)\big]=k(X_0)+ \int_0^t \mathbb{E}\mathcal{A}k(X_u)\po u.
\end{equation}
This will be used later in a statistical procedure.
Our first main result identifies the generator $\mathcal{A}$.

\begin{theorem}\label{theorem:generator}
Functions $k(\lambda,\,n)$ of the form $k(\lambda,\,n)=g(\lambda)f(n)$, where $f$ is a measurable function on $\mathbb{N}\cup\{0\}$
and $g$ is a function continuously differentiable on nonnegative half-line, belong to the domain of the $\mathcal{D}(\mathcal{A})$ of generator $\mathcal{A}$. Further,
\begin{equation*}
    (\mathcal{A}k)(\lambda_t, N_t)=\mathcal{A}\big[g(\lambda_t)f(N_t)\big] = \lambda_t\,\big[f(N_t+1)g(\lambda_t+\alpha)-f(N_t)g(\lambda_t)\big]+ f(N_t)\,g'(\lambda_t)\,\beta\,(\lambda_\infty-\lambda_t)\,.
\end{equation*}
\end{theorem}
The proof of this theorem will be given in Appendix.

\section{Parameters estimation}\label{chap3}
We will now present the moment method of estimating all parameters of Hawkes process, that is, estimating
$\lambda_\infty $, $\lambda_0$, $\beta$ and $\alpha$.
We observe our Hawkes process within finite time, that is between $t$ and $t+\Delta$, where $\Delta>0$.
We will identify above parameters by finding moments
$ \mathbb{E}\big[N_{t+\Delta}-N_t\big]^k$.

The main idea is based on using formula \eqref{keyidenity} together with Theorem \ref{theorem:generator}.
Taking $k(X_t)=k(\lambda_t, N_t)=N_t^l\,(l\in\mathbb{N})$ we get
\begin{equation}\label{Nl}
    \mathbb{E}[N_t^l] = \mathbb{E}\int_0^t \lambda_u\big((N_u +1)^l-N_u^l\big)\po u
\end{equation}
which is equivalent to
\begin{equation*}
    \mathbb{E}[N_t^l] = \int_0^t\sum_{k=0}^{l-1}\binom{l}{k}\,\mathbb{E}[\lambda_u N_u^k]\po u = \sum_{k=0}^{l-1}\binom{l}{k}\,\int_0^t\mathbb{E}[\lambda_u N_u^k]\po u.
\end{equation*}
Hence
\begin{equation}\label{f1}
    \frac{\po}{\po t}\mathbb{E}[N_t^l] = \sum_{k=0}^{l-1}\binom{l}{k}\,\mathbb{E}[\lambda_t N_t^k].
\end{equation}
Similarly, by taking $k(X_t)=k(\lambda_t, N_t)=\lambda_t^m$ and $k(X_t)=k(\lambda_t, N_t)=\lambda_t^m N_t^l$, we derive
\begin{equation}\label{f2}
    \frac{\po}{\po t}\mathbb{E}[\lambda_t^m] =
    m\beta\lambda_\infty\mathbb{E}[\lambda_t^{m-1}]-m\beta\mathbb{E}[\lambda_t^m]+\sum_{j=0}^{m-1}\binom{m}{j}\,\alpha^{m-j}\mathbb{E}[\lambda_t^{j+1}]
\end{equation}
and
\begin{equation}\label{f3}
\begin{split}
    \frac{\po}{\po t}\mathbb{E}[\lambda_t^m\,N_t^l] & =
    m\beta\lambda_\infty\mathbb{E}[\lambda_t^{m-1} N_t^l] - m\beta\mathbb{E}[\lambda_t^m N_t^l]\,+ \\
    & + \sum_{j=0}^m\sum_{k=0}^l\binom{m}{j}\binom{l}{k}\,\alpha^{m-j}\mathbb{E}[\lambda_t^{j+1}N_t^k] - \mathbb{E}[\lambda_t^{m+1}N_t^l].
\end{split}
\end{equation}
These identities will produce equations for the mixed moments $\mathbb{E}[\lambda_t^m\,N_t^l]$ as functions of $t$
that could be solved explicitly. These moments allow to
find $ \mathbb{E}\big[N_{t+\Delta}-N_t\big]^k$ for $k=1,2,3$.
We start first from finding the first moment given in the following theorem.
We denote
\begin{equation}\label{lambdastar}
\lambda^*=\frac{\beta\lambda_\infty}{\beta-\alpha}.\end{equation}
\begin{theorem}\label{theorem:limEN}
We have
\begin{equation}\label{truefirstmoment}
    \mathbb{E}\big[N_{t+\Delta}-N_t\big] = \lambda^*\Delta + \frac{\lambda_0-\lambda^*}{\beta-\alpha}(e^{-(\beta-\alpha)t}-e^{-(\beta-\alpha)(t+\Delta)}).
\end{equation}
and
    \begin{equation}\label{limitmomemnt}
    \lim_{t\to\infty}\mathbb{E}\big[N_{t+\Delta}-N_t\big] = \lambda^*\Delta = \frac{\beta\lambda_\infty}{\beta-\alpha}\,\Delta.
    \end{equation}
\end{theorem}
The proof of this theorem will be given in Appendix.
The expression given in \eqref{truefirstmoment}
depends on the initial intensity $\lambda_0$, but based on the equation~(\ref{eq:lambda_wyprowadzony_wzor}) we can conclude that for increasingly larger $t$ the influence of this constant on the value of the process intensity function is negligible.
Therefore following~\cite{financial_contagion} we focus on the long-run behavior of the process from now on.
We need then two more moments to find $3$ parameters of Hawkes process, namely, $\lambda_\infty $, $\beta$ and $\alpha$.

Later, in Appendix we will prove another two key facts given in next theorem.
\begin{theorem}\label{theorem:limEN2}
   We have
    \begin{equation}\label{firsttheorem}
    \begin{split}
    \lim_{t\to\infty}\mathbb{E}\big[(N_{t+\Delta}-N_t)^2\big] = \frac{\beta\lambda_\infty}{(\beta-\alpha)^4} & \Big[\alpha(2\beta-\alpha)\,e^{-(\beta-\alpha)\Delta}+\alpha(\alpha-2\beta)\\
    & + \Delta\beta^2(\beta-\alpha)+\Delta^2\beta\lambda_\infty(\beta-\alpha)^2\Big]
    \end{split}
    \end{equation}
    and
    \begin{equation}\label{secondtheorem}
    \begin{split}
    \lim_{t\to\infty}\mathbb{E}\big[(N_{t+\Delta}-N_t)^3\big] & = \Delta^3\frac{\beta^3\lambda_\infty^3}{(\beta-\alpha)^3} + \Delta^2\frac{3\beta^4\lambda_\infty^2}{(\beta-\alpha)^4}\\
    & + \Delta\frac{\beta^2\lambda_\infty}{(\beta-\alpha)^5}\big(3\lambda_\infty\alpha(\alpha-2\beta)+\beta^2(2\alpha+\beta)\big)\\
    & + \frac{3\alpha\beta^2\lambda_\infty}{2(\beta-\alpha)^6}\big(\alpha^2-\alpha\beta-4\beta^2\big) + \frac{\alpha^2\beta\lambda_\infty(2\alpha-3\beta)}{2(\beta-\alpha)^5}\,e^{-2(\beta-\alpha)\Delta}\\
    & + \frac{\alpha\beta\lambda_\infty}{(\beta-\alpha)^6}\big(\alpha^3-4\alpha^2\beta+3\alpha\beta^2+6\beta^3\big)\,e^{-(\beta-\alpha)\Delta}\\
    & - \frac{3\alpha\beta^2\lambda_\infty(\lambda_\infty+\alpha)(\alpha-2\beta)}{(\beta-\alpha)^5}\,\Delta e^{-(\beta-\alpha)\Delta}.
    \end{split}
    \end{equation}
\end{theorem}

Having moments $ \mathbb{E}\big[N_{t+\Delta}-N_t\big]^k$ for $k=1,2,3$ identified we can produce moment estimators
$\hat{\alpha}$, $\hat{\beta}$ and $\hat{\lambda}_\infty$ of $\alpha$, $\beta$ and $\lambda_\infty$, respectively, by
solving the system of equations:
\begin{equation}\label{system}
    \begin{cases}
        \mathbb{M}_1 & = \lim_{t\to\infty}\mathbb{E}[N_{t+\Delta}-N_{t}] \\
        \mathbb{M}_2 & = \lim_{t\to\infty}\mathbb{E}[(N_{t+\Delta}-N_{t})^2] \\
        \mathbb{M}_3 & = \lim_{t\to\infty}\mathbb{E}[(N_{t+\Delta}-N_{t})^3]
    \end{cases},
\end{equation}
where $\mathbb{M}_i$ is the arithmetic mean of the expressions of the form $(N_{t_j+\Delta}-N_{t_j})^i$, $t_j$ are the points in time mutually distant from each other by $\Delta$ and $N_{t_j}$ are observed values of our process.

To analyze how good is this estimation method we generated Hawkes process' trajectories using
the algorithm mentioned in Section \ref{chap1} (see:~Appendix, Algorithm \ref{kod}).
We set the parameters to $\alpha=0.2$, $\beta=1$ and~$\lambda_\infty=1$. Due to the fact that the derived formulas are true for large $t$, it was decided to take into account the values of $N_{t+\Delta}-N_t$ for $t$ between $3000$ and $10000$. Besides it, we took $\Delta=0.5$.
The system of equation \eqref{system}
is non-linear, and the function \textit{nleqslv} built in one of the \textit{R} programming language packages was used to solve it. As many other numerical methods, it was necessary to specify the initial values for the algorithm which were set to $\alpha=0.5$, $\beta=1.5$ and~$\lambda_\infty=2$.

\begin{table}[ht]
\centering
\begin{tabular}{|c|c|c|c||c|c|c|c|}
\hline
Number of simulation & $\hat{\alpha}$ & $\hat{\beta}$ & $\hat{\lambda}_\infty$ & Number of simulation & $\hat{\alpha}$ & $\hat{\beta}$ & $\hat{\lambda}_\infty$\\
\hline\hline
1 & 0.192 & 0.877 & 0.988 & 11 & 0.191 & 1.432 & 1.086 \\
2 & 0.220 & 2.153 & 1.126 & 12 & 0.221 & 1.342 & 1.065 \\
3 & 0.164 & 0.681 & 0.960 & 13 & 0.177 & 1.085 & 1.013 \\
4 & 0.235 & 1.376 & 1.026 & 14 & 0.187 & 2.100 & 1.127 \\
5 & 0.119 & 0.673 & 1.007 & 15 & 0.233 & 1.809 & 1.101 \\
6 & 0.180 & 0.915 & 1.009 & 16 & 0.212 & 1.503 & 1.089 \\
7 & 0.200 & 0.967 & 1.000 & 17 & 0.241 & 1.418 & 1.053 \\
8 & 0.189 & 2.228 & 1.155 & 18 & 0.185 & 1.456 & 1.091 \\
9 & 0.197 & 1.067 & 1.032 & 19 & 0.202 & 1.391 & 1.089 \\
10 & 0.251 & 1.713 & 1.090 & 20 & 0.161 & 0.816 & 1.009 \\
\hline
\end{tabular}
\caption{Estimation results for real parameter values $\alpha = 0.2,\,\beta=1,\,\lambda_\infty = 1$.}
\label{tab:symulacje}
\end{table}

\begin{figure}[h]
    \centering
    \includegraphics[scale=1]{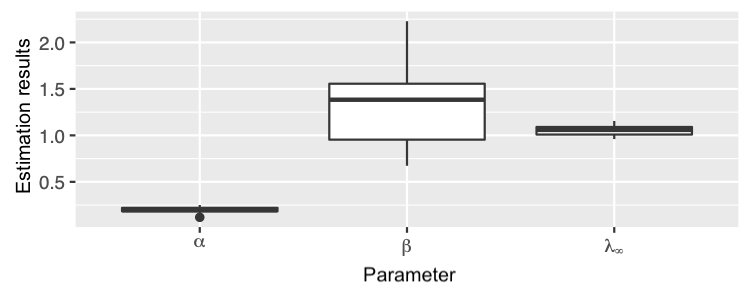}
    \caption{Boxplot of estimation results for 20 process trajectories.}
    \label{fig:boxploty}
\end{figure}

The results of the estimation are given in Table \ref{tab:symulacje}. Notice that the estimator of the parameter $\lambda_\infty$ fares surprisingly well. Based on twenty simulations, we obtain an average result of $1.056\pm0.054$, which is very close to the real value $\lambda_\infty=1$. For the estimators of $\alpha$ and $\beta$ , we get average values of $0.198\pm0.031$ (which is also a very good result) and $1.350\pm0.474$, respectively.  Figure~\ref{fig:boxploty} shows that despite the identical values of the actual $\beta$ and $\lambda_\infty$, as well as the fact that the initial $\beta$ value given for the algorithm was closer to the real value than it was for $\lambda_\infty$, the $\hat{\beta}$ estimator is much worse than $\hat{\lambda}_\infty$.

\section{Twitter data analysis}\label{sectwitter}
Our last and main goal is to show that Hawkes process can indeed model
clustering behaviour in social media.
We decided to analyze one of the most popular websites, Twitter, which allows users to re-share content posted by other people with public accounts. This is called a retweet. With each retweet, the original post reaches a new audience who can also pass it on to their followers. If we marked on the axis the time points at which users retweeted the original tweet, we would most likely observe their clustering.

\begin{figure}[h]
    \centering
    \includegraphics[width=0.6\textwidth]{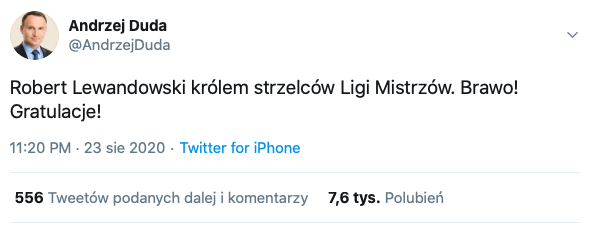}
    \caption{Analyzed tweet from the Polish president's account.}
    \label{fig:tweet}
\end{figure}

One of the most popular accounts on Polish Twitter is the account of the president of Poland, Andrzej Duda (\textit{@AndrzejDuda}). Due to its popularity, it may prove to be an interesting source of data for analysis and modeling using the Hawkes process. This account is public, so each tweet can be retweeted by other Twitter users. The post presented in Fig. \ref{fig:tweet} was retweeted exactly $556$ times at the time of analysis.

\begin{figure}[h]
    \centering
    \includegraphics[width=0.7\textwidth]{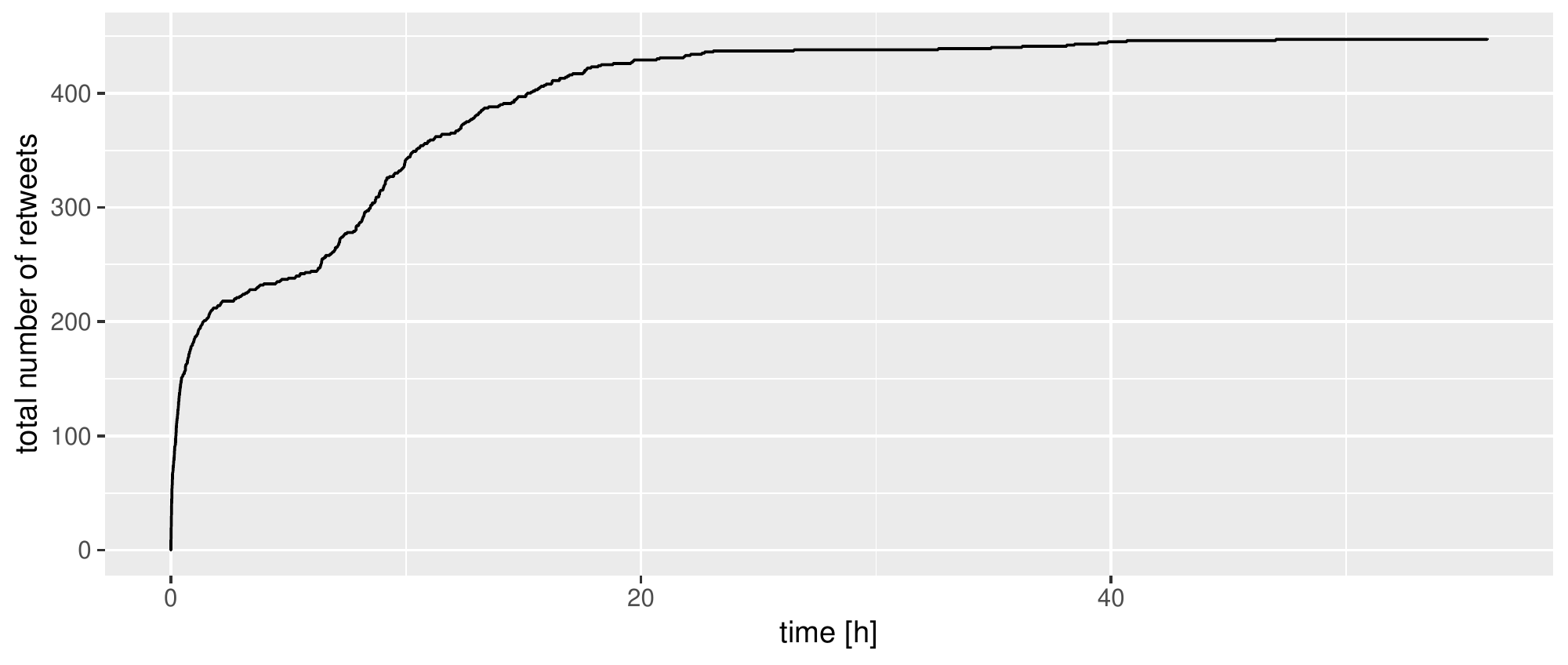}
    \caption{Total number of retweets over time.}
    \label{fig:retweets_in_time}
\end{figure}

Unfortunately, due to Twitter API limitations and user privacy settings, it is not possible to obtain accurate data. Some users block access to their tweets for people who do not follow them, so we cannot obtain information about the time when they retweeted the post we were analyzing. Using the \textit{Python tweepy} module, which enables easy use of Twitter API, it was possible to obtain information on $448$ retweets, which is about 80\% of the total. Their distribution over time is shown in the graph in Figure \ref{fig:retweets_in_time}, where the horizontal axis represents the time since the original post was published.

\begin{figure}[h]
    \centering
    \includegraphics[width=0.7\textwidth]{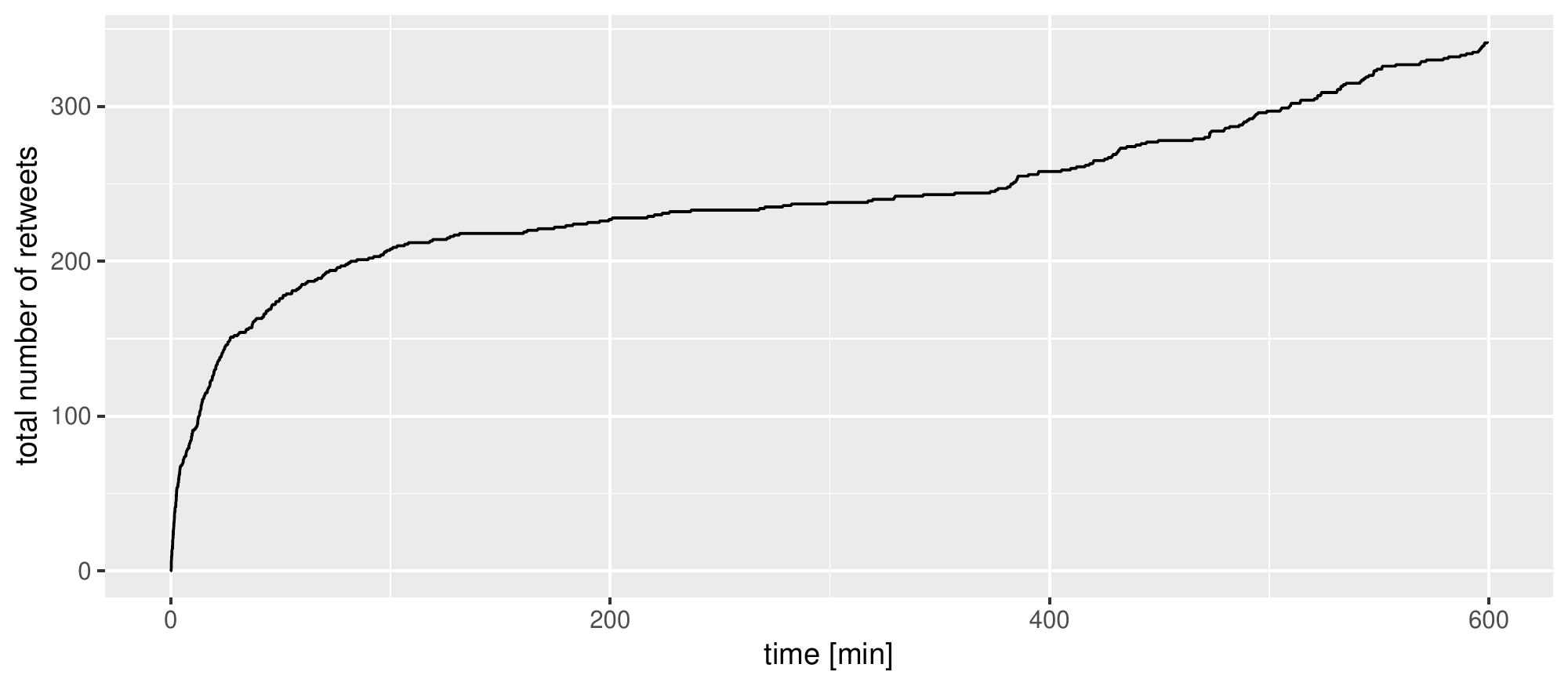}
    \caption{Number of retweets on the limited time horizon.}
    \label{fig:retweets_cut}
\end{figure}

There are clear changes in dynamics here. The total number of retweets increases sharply in the first few minutes after publishing the post. Then the speed of growth slows down slightly to rise again after about $6$ hours (perhaps the post was retweeted by another popular account). Further, the process stabilizes, the increase in the number of retweets is negligible. In \cite{hawkes_in_social_media}, a similar analysis was performed, in which the author of the post was an account with over $12,000$ followers, and the time horizon was set to $600$ seconds. Since Andrzej Duda's account has definitely more followers (over $1.1$ million), adopting a time horizon of $10$ hours seems to make sense. The time for a tweet to be retweeted by another user is made available with an accuracy of one second, but because the delta should be small, the base unit we take is minutes, so $\Delta=\frac{1}{60}\,[min]$, and the time horizon was set at $10$ hours. Figure \ref{fig:retweets_cut} presents the analyzed data.

\begin{figure}[h]
    \centering
    \includegraphics[width=0.7\textwidth]{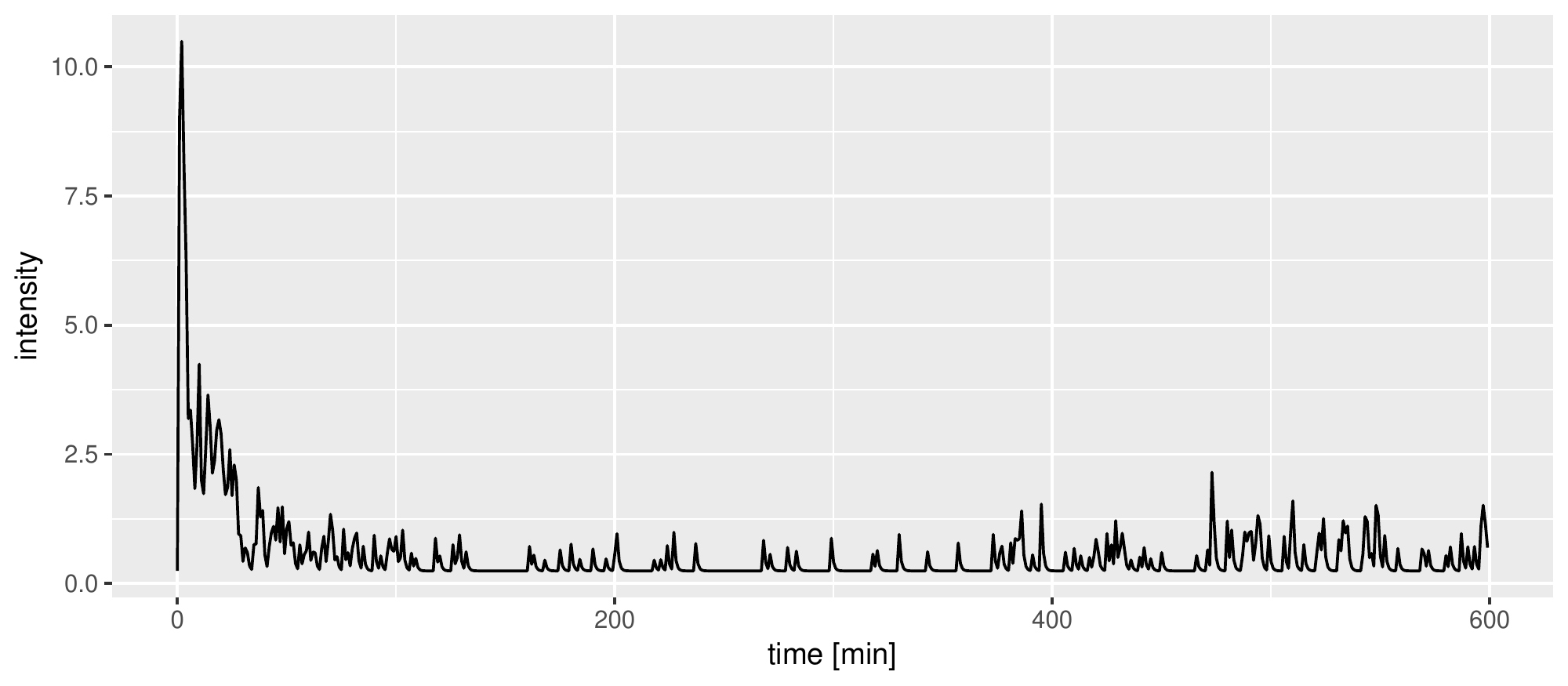}
    \caption{Process intensity graph for parameters $\alpha = 0.772, \beta=1.133, \lambda_\infty = \lambda_0 = 0.243$.}
    \label{fig:intensity}
\end{figure}

The numerical method of solving a system of nonlinear equations \eqref{system} requires specifying an initial point, which can be any of the domains we allow. Assuming $(\alpha, \beta, \lambda_\infty) = (0.5, 1.5, 0.75)$ as the starting point, the result of the estimation is approximately $\alpha = 0.772, \beta=1.133, \lambda_\infty = 0.243$. Assuming that $\lambda_0$ is the same as $\lambda_\infty$, , the intensity graph of this process would look like in Fig. \ref{fig:intensity}. Using the formula from equation \ref{eq:wartosc_oczekiwana_Nt} we can check that with these parameters the expected value of the process at time $t=600$ (10 hours) is $\mathbb{E}N_{600}=348.7265$. The actual number of retweets up to that point was $342$.

\begin{figure}[h]
    \centering
    \includegraphics[width=0.7\textwidth]{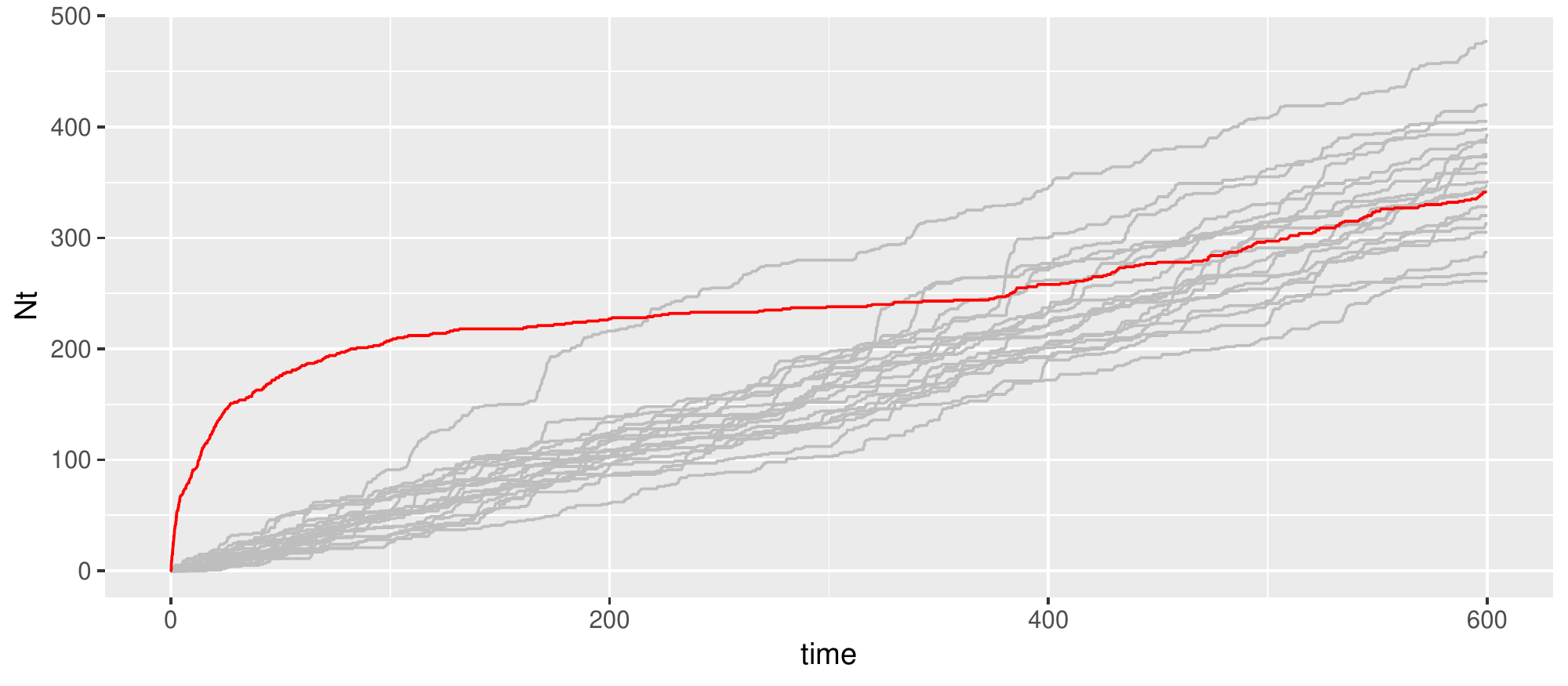}
    \caption{Comparison of real data with simulation results.}
    \label{fig:compared_data}
\end{figure}

To validate the estimation results, twenty possible Hawkes process trajectories were generated for the parameter set $\alpha = 0.772, \beta=1.133, \lambda_\infty = \lambda_0 = 0.243$. These trajectories, marked with gray lines, were compared in Fig. \ref{fig:compared_data} with the actual number of retweets in time (marked with a red line). We observe a significant deviation from the norm at the beginning of the process, the number of retweets in the first minutes increases too rapidly. It only starts to resemble the generated trajectories after about $400$ minutes.
This may suggest that a different model would be necessary to be able to describe the dynamics of the spread of content posted by popular accounts in the early stage of the process.

\section{Conclusions}\label{secconcl}
In this paper, using the generator theory, we describe moments method of estimation of the parameters of Hawkes process.
We apply it to show that in social media quite often the clustering phenomenon might appear.
As an example we analyze retweet cascades.
The results based on the twenty generated trajectories turned out to be surprisingly good, however, showing a relatively large variance of the $\beta$ parameter estimate compared to the other two.
It would be good to look for other possible applications of Hawkes process in
social media and present paper can seen as the first step toward
deeper understanding the contagion dynamics there.

\section{Appendix}
\subsection{Proof of Theorem \ref{theorem:generator}}
Observe that
\begin{equation}\label{0lim}
\begin{split}
    \mathcal{A}\,g(\lambda_t)f(N_t) & = \lim_{h\to0}\frac{1}{h}\mathbb{E}\big[g(\lambda_{t+h})f(N_{t+h})-g(\lambda_t)f(N_t)\,|\,\mathcal{F}_t\big] = \\
    & = \lim_{h\to0}\frac{1}{h}\mathbb{E}\big[\big(f(N_{t+h})-f(N_t)\big)\big(g(\lambda_{t+h})-g(\lambda_t)\big)\,|\,\mathcal{F}_t\big] \\
    & + f(N_t)\lim_{h\to0}\frac{1}{h}\mathbb{E}\big[g(\lambda_{t+h})-g(\lambda_t)\,|\,\mathcal{F}_t\big] + g(\lambda_t)\lim_{h\to0}\frac{1}{h}\mathbb{E}\big[f(N_{t+h})-f(N_t)\,|\,\mathcal{F}_t\big].
\end{split}
\end{equation}
Note that the limits that appear in the last line are equal to $\mathcal{A}^\lambda g$ and $\mathcal{A}^N f$, respectively, being the generators of $\lambda_t$ and~$N_t$ applied to the functions $g$ and~$f$.
More formally,
\begin{equation}\label{1lim}
    \mathcal{A}^N f(N_t) = \lim_{h\to0}\frac{1}{h}\mathbb{E}\big[f(N_{t+h})-f(N_t)\,|\,\mathcal{F}_t\big]
    =\lambda_t\big[f(N_t+1)-f(N_t)\big]
\end{equation}
and by \eqref{eq:d_lambda}
\begin{equation}\label{2lim}
\begin{split}
    \mathcal{A}^\lambda g(\lambda_t) & = \lim_{h\to0}\frac{1}{h}\mathbb{E}\big[g(\lambda_{t+h})-g(\lambda_t)\,|\,\mathcal{F}_t\big] = \\
    & = \lim_{h\to0}\frac{1}{h}\mathbb{E}\big[g(\lambda_t+\beta h(\lambda_\infty-\lambda_t)+\alpha(N_{t+h}-N_t))-g(\lambda_t)\,|\,\mathcal{F}_t\big]\\
    &= \lim_{h\to0}\frac{1}{h}\mathbb{E}\big[g(\lambda_{t+h})-g(\lambda_t)\,|\,\mathcal{F}_t\big] = \\
    & = \lim_{h\to0}\frac{1}{h}\mathbb{E}\big[g(\lambda_t+\beta h(\lambda_\infty-\lambda_t)+\alpha(N_{t+h}-N_t))-g(\lambda_t)\,|\,\mathcal{F}_t\big]\\
    &=\lambda_t\big[g(\lambda_t+\alpha)-g(\lambda_t)\big] + \beta(\lambda_\infty-\lambda_t)g'(\lambda_t)
\end{split}
\end{equation}
where we used fact that
\begin{equation*}
    \lim_{h\to0}\frac{g(\lambda_t+\beta h(\lambda_\infty-\lambda_t))-g(\lambda_t)}{h} = \beta(\lambda_\infty-\lambda_t)g'(\lambda_t).
\end{equation*}
Furthermore,
\begin{equation}\label{3lim}
\begin{split}
    \lim_{h\to0}&\,\frac{1}{h}\, \mathbb{E}\big[\big(f(N_{t+h})-f(N_t)\big)\big(g(\lambda_{t+h})-g(\lambda_t)\big)\,|\,\mathcal{F}_t\big] = \\
    & = \lim_{h\to0}\frac{1}{h}\,\mathbb{E}\big[\big(f(N_{t+h})-f(N_t)\big)\big(g(\lambda_t+\beta h(\lambda_\infty-\lambda_t)+\alpha(N_{t+h}-N_t))-g(\lambda_t)\big)\,|\,\mathcal{F}_t\big] = \\
    & = \big(f(N_t+1)-f(N_t)\big)\big(g(\lambda_t+\alpha)-g(\lambda_t)\big)\,\lambda_t.
\end{split}
\end{equation}
Putting the limits \eqref{1lim}, \eqref{2lim} and \eqref{3lim} into \eqref{0lim} completes the proof.
\QEDA
\vspace{0.8cm}

\subsection{Proof of Theorem \ref{theorem:limEN}}
From equations (\ref{eq:lambda_wyprowadzony_wzor}), (\ref{eq:dNt}) and Fubini's theorem we have
\begin{equation*}
    \mu(t)= \lambda_\infty + e^{-\beta t}(\lambda_0-\lambda_\infty) + \alpha\int_0^t e^{-\beta(t-s)}\mu(s)\po s,
\end{equation*}
where $\mu(t)=\mathbb{E}\lambda_t$
which after differentiation over time simplifies to the standard non-homogeneous differential equation:
\begin{equation*}
    \frac{\po\mu}{\po t}=\beta\lambda_\infty-(\beta-\alpha)\mu
\end{equation*}
with the initial condition $\mu(0)=\mathbb{E}\lambda_0=\lambda_0$. Thus
\begin{equation*}
    \mu(t)=\mathbb{E}\lambda_t=\lambda^*-(\lambda^*-\lambda_0)e^{-(\beta-\alpha)t},
\end{equation*}
where $\lambda^*$ is defined in \eqref{lambdastar}.
Since by \eqref{Nl},
\begin{equation}\label{Ent}
    \int_0^t\mathbb{E}\lambda_s\po s =  \mathbb{E}N_t
\end{equation}
we derive that equation \eqref{truefirstmoment}. Equation \eqref{limitmomemnt} is straightforward from it.
\QEDA
\vspace{0.8cm}

\subsection{Proof of Theorem \ref{theorem:limEN2}}
We first prove identity \eqref{firsttheorem}.
Using Theorem \ref{theorem:generator} with
$k(\lambda,\,n)=n$ and $k(\lambda,\,n)=n^2$ we can write
\begin{equation*}
    \mathcal{A}\,N_t = \lambda_t\quad\text{and}\quad\mathcal{A}\,N_t^2 = 2\lambda_t N_t+\lambda_t.
\end{equation*}
Thus from \eqref{keyidenity}
\begin{equation*}
    \mathbb{E}\big(N_{t+\Delta}\,|\,\mathcal{F}_t\big) = N_t + \int_t^{t+\Delta}\mathbb{E}\big(\lambda_u\,|\,\mathcal{F}_t\big)\po u
\end{equation*}
and
\begin{equation*}
    \mathbb{E}\big(N_{t+\Delta}^2\,|\,\mathcal{F}_t\big) = N_t^2 + 2\int_t^{t+\Delta}\mathbb{E}\big(\lambda_u N_u\,|\,\mathcal{F}_t\big)\po u + \int_t^{t+\Delta}\mathbb{E}\big(\lambda_u\,|\,\mathcal{F}_t\big)\po u.
\end{equation*}
Denoting $\Delta N_t = N_{t+\Delta}-N_t$, we can conclude that
\begin{equation}\label{trans1}
    \mathbb{E}\big[(\Delta N_t)^2\big] = 2\int_t^{t+\Delta}\mathbb{E}(\lambda_u N_u)\po u + \int_t^{t+\Delta}\mathbb{E}(\lambda_u)\po u - 2\mathbb{E}\Big(N_t\,\int_t^{t+\Delta}\mathbb{E}\big(\lambda_u\,|\,\mathcal{F}_t\big)\po u\Big).
\end{equation}
Using identity \eqref{f2} we can determine $\mathbb{E}\lambda_t^2$ by solving differential equation
\begin{equation*}
    \frac{\po}{\po t}\mathbb{E}\lambda_t^2 = (\alpha^2+2\beta\lambda_\infty)\mathbb{E}\lambda_t -2(\beta-\alpha)\mathbb{E}\lambda_t^2.
\end{equation*}
producing
\begin{equation*}
\begin{split}
    \mathbb{E}\lambda_t^2 = & (\lambda^*)^2 + \frac{\alpha^2\lambda^*}{2(\beta-\alpha)} + \frac{(\lambda_0-\lambda^*)(\alpha^2 + 2\beta\lambda_{\infty})}{\beta-\alpha}\,e^{-(\beta-\alpha)t} \\
    + & \left((\lambda_0-\lambda^*)^2 - \frac{\alpha^2(2\lambda_0-\lambda^*)}{2(\beta-\alpha)}\right)\,e^{-2(\beta-\alpha)t}.
\end{split}
\end{equation*}
Moreover from \eqref{f3}
the expression inside the first integral in \eqref{trans1} solves the following differential equation
\begin{equation}\label{eq:d_lambda_N}
    \frac{\po}{\po t}\mathbb{E}(\lambda_t N_t) = \beta\lambda_\infty\mathbb{E}N_t+\mathbb{E}\lambda_t^2+\alpha\mathbb{E}\lambda_t-(\beta-\alpha)\mathbb{E}(\lambda_t N_t)
\end{equation}
where $\mathbb{E}N_t$ is given in \eqref{Ent}. Hence the first integral in \eqref{trans1} equals
\begin{equation}\label{eq:pierwsza_calka}
\begin{split}
    \int_t^{t+\Delta}\mathbb{E}(\lambda_u N_u)\po u & = \mathbb{E}(\lambda_t N_t) \int_t^{t+\Delta} e^{-(\beta-\alpha)(u-t)}\po u \\
    & + \beta\lambda_\infty\mathbb{E}N_t \int_t^{t+\Delta}\int_t^u e^{-(\beta-\alpha)(u-s)}\po s\po u \\
    & + \int_t^{t+\Delta}\int_t^u e^{-(\beta-\alpha)(u-s)}\big[\beta\lambda_\infty\int_t^s \mathbb{E}\lambda_r\po r+\mathbb{E}\lambda_s^2+\alpha\mathbb{E}\lambda_s \big]\po s\po u.
\end{split}
\end{equation}
To find last integral in \eqref{trans1} we take $k(\lambda,n)=\lambda$
in Theorem~\ref{theorem:generator} and in \eqref{keyidenity} to derive
for $u\geq t$,
\begin{equation*}
    \po\mathbb{E}\big(\lambda_u\,|\,\mathcal{F}_t\big) = \beta\lambda_\infty\po u -(\beta-\alpha)\mathbb{E}\big(\lambda_u\,|\,\mathcal{F}_t\big)\po u,
\end{equation*}
which gives
\begin{equation}\label{eq:trzecia_calka}
\begin{split}
    \mathbb{E}\Big(N_t\,\int_t^{t+\Delta}\mathbb{E}\big(\lambda_u\,|\,\mathcal{F}_t\big)\po u\Big) & = \mathbb{E}(\lambda_t N_t)\int_t^{t+\Delta} e^{-(\beta-\alpha)(u-t)}\po u\\
    & + \beta\lambda_\infty\mathbb{E}N_t \int_t^{t+\Delta}\int_t^u e^{-(\beta-\alpha)(u-s)}\po s\po u.
\end{split}
\end{equation}
Keeping in mind that $\mathbb{E}\lambda_t$ and~$\mathbb{E}\lambda_t^2$ are known to us, we can write the following explicit formula:
\begin{equation*}
\begin{split}
    \mathbb{E}\big[(\Delta N_t)^2\big] & =  \int_t^{t+\Delta}\mathbb{E}(\lambda_u)\po u \\
    & + 2\int_t^{t+\Delta}\int_t^u e^{-(\beta-\alpha)(u-s)} \big[\beta\lambda_\infty\int_t^s \mathbb{E}\lambda_r\po r + \mathbb{E}\lambda_s^2+\alpha\mathbb{E}\lambda_s\big]\po s\po u;
\end{split}
\end{equation*}
see also~\cite{fonseca_hawkes_definition} for similar considerations.
The first and second moments of $\lambda_t$ are known to us, and their limit values are as follows:
\begin{equation*}
    \Lambda_1 = \lim_{t\to\infty}\mathbb{E}\lambda_t = \frac{\beta\lambda_\infty}{\beta-\alpha} \quad\text{and}\quad \Lambda_2 = \lim_{t\to\infty}\mathbb{E}\lambda_t^2 = \frac{\beta\lambda_\infty(\alpha^2+2\beta\lambda_\infty)}{2(\beta-\alpha)^2}\,.
\end{equation*}
Thus
\begin{equation}\label{eq:N2_dowstawienia}
    \lim_{t\to\infty}\mathbb{E}\big[(\Delta N_t)^2\big] =  \Lambda_1\Delta +
    2\beta\lambda_\infty\Lambda_1\,I_1 + 2\big[ \Lambda_2+\alpha\Lambda_1\big]\,I_2,
\end{equation}
where
\begin{equation*}
    I_1 = \int_t^{t+\Delta}\int_t^u\int_t^s e^{-(\beta-\alpha)(u-s)}\po r\po s\po u = \frac{\Delta^2}{2(\beta-\alpha)}-\frac{\Delta}{(\beta-\alpha)^2}-\frac{1}{(\beta-\alpha)^3}\big(e^{-(\beta-\alpha)\Delta}-1\big)
\end{equation*}
and
\begin{equation*}
    I_2 = \int_t^{t+\Delta}\int_t^u e^{-(\beta-\alpha)(u-s)}\po s\po u = \frac{\Delta}{\beta-\alpha}+\frac{1}{(\beta-\alpha)^2}\big(e^{-(\beta-\alpha)\Delta}-1\big).
\end{equation*}
This completes the proof of \eqref{firsttheorem}.

To show \eqref{secondtheorem}
we denote $M_3=\mathbb{E}\big[(\Delta N_t)^3\big]$.
In next step we use \eqref{f2}
to produce equation for $\mathbb{E}\lambda_t^3$. By solving it and by taking limit with $t\rightarrow \infty$ we derive
\begin{equation*}
    \Lambda_3 = \lim_{t\to\infty}\mathbb{E}\lambda_t^3 = \frac{\alpha^3\beta\lambda_\infty}{3(\beta-\alpha)^2} + \frac{\beta\lambda_\infty(\alpha^2+\beta\lambda_\infty)(\alpha^2+2\beta\lambda_\infty)}{2(\beta-\alpha)^3}
\end{equation*}
Hence
\begin{equation*}
    M_3=\mathbb{E}\big[(\Delta N_t)^3\big] = \mathbb{E}\big[\mathbb{E}[N_{t+\Delta}^3\,|\,\mathcal{F}_t]-3 N_t\mathbb{E}[N_{t+\Delta}^2\,|\,\mathcal{F}_t]+3 N_t^2\mathbb{E}[N_{t+\Delta}\,|\,\mathcal{F}_t]-N_t^3\big].
\end{equation*}
Taking $k(\lambda,\,n)=n^i$ with $i=1,2,3$ in Theorem~\ref{theorem:generator} and in \eqref{keyidenity} we get
\begin{equation*}
\begin{split}
    M_3 & = \int_t^{t+\Delta}\mathbb{E}[\lambda_s+3\lambda_s N_s+3\lambda_s N_s^2]\po s - 3\,\mathbb{E}\big[N_t \int_t^{t+\Delta}\mathbb{E}[\lambda_s+2\lambda_s N_s\,|\,\mathcal{F}_t]\po s\big]\\
    & + 3\,\mathbb{E}\big[N_t^2 \int_t^{t+\Delta}\mathbb{E}[\lambda_s\,|\,\mathcal{F}_t]\po s\big]
\end{split}
\end{equation*}
which is equivalent by similar considerations as before to
\begin{equation*}
\begin{split}
    M_3 & = \int_t^{t+\Delta}\mathbb{E}\lambda_u\po u + 3\int_t^{t+\Delta}\int_t^u e^{-(\beta-\alpha)(u-s)}\big[\beta\lambda_\infty\int_t^s \mathbb{E}\lambda_r\po r+\mathbb{E}\lambda_s^2+\alpha\mathbb{E}\lambda_s\big]\po s\po u \\
    & + 3 \int_t^{t+\Delta}\mathbb{E}(\lambda_u N_u^2)\po u - 6\mathbb{E}\big[N_t \int_t^{t+\Delta}\mathbb{E}[\lambda_u N_u\,|\,\mathcal{F}_t]\po u\big] + 3\mathbb{E}\big[N_t^2 \int_t^{t+\Delta}\mathbb{E}[\lambda_u\,|\,\mathcal{F}_t]\po u\big].
\end{split}
\end{equation*}
Further analysis shows that
\begin{equation*}
\begin{split}
\mathbb{E}[N_{s}\,|\,\mathcal{F}_t]&=N_t+\int_t^s\mathbb{E}[\lambda_r\,|\,\mathcal{F}_t]\po r,\\
    \mathbb{E}(\lambda_u N_u^2) & = \mathbb{E}(\lambda_t N_t^2)\,e^{-(\beta-\alpha)(u-t)} \\
    & + \int_t^u e^{-(\beta-\alpha)(u-s)}[\mathbb{E}\lambda_s^2+2\mathbb{E}(\lambda^2_s N_s)+\alpha\mathbb{E}\lambda_s+2\alpha\mathbb{E}(\lambda_s N_s)+\beta\lambda_\infty\mathbb{E}N_s^2]\po s,\\
    \mathbb{E}[N_s^2]&=\mathbb{E}N_t^2+\int_t^s\mathbb{E}[\lambda_r+2\lambda_r N_r]\po r
\end{split}
\end{equation*}
which together with \eqref{eq:pierwsza_calka} gives
\begin{equation*}
\begin{split}
    M_3 & = \int_t^{t+\Delta}\mathbb{E}\lambda_u\po u + 6\int_t^{t+\Delta}\int_t^u e^{-(\beta-\alpha)(u-s)}\big[\beta\lambda_\infty\int_t^s \mathbb{E}\lambda_r\po r+\mathbb{E}\lambda_s^2+\alpha\mathbb{E}\lambda_s\big]\po s\po u\\
    & + 6\int_t^{t+\Delta}\int_t^u e^{-(\beta-\alpha)(u-s)}[\mathbb{E}(\lambda^2_s N_s)+\alpha\mathbb{E}(\lambda_s N_s)]\po s\po u\\
    & - 6\mathbb{E}\big[N_t \int_t^{t+\Delta}\int_t^u e^{-(\beta-\alpha)(u-s)}\mathbb{E}[\lambda_s^2+\alpha\lambda_s\,|\,\mathcal{F}_t]\po s\po u\big] \\
    & + 6\beta\lambda_\infty\int_t^{t+\Delta}\int_t^u\int_t^s e^{-(\beta-\alpha)(u-s)}\mathbb{E}(\lambda_r N_r)\po r\po s\po u\\
    & - 6\beta\lambda_\infty\mathbb{E}\big[N_t \int_t^{t+\Delta}\int_t^u\int_t^s e^{-(\beta-\alpha)(u-s)}\mathbb{E}[\lambda_r\,|\,\mathcal{F}_t]\po r\po s\po u\big].
\end{split}
\end{equation*}
Using the previously derived formulas, some components can be reduced, to produce the final
expression:
\begin{equation*}
\begin{split}
    M_3 & = \int_t^{t+\Delta}\mathbb{E}\lambda_u\po u + 6\int_t^{t+\Delta}\int_t^u e^{-(\beta-\alpha)(u-s)}\big[\beta\lambda_\infty\int_t^s \mathbb{E}\lambda_r\po r+\mathbb{E}\lambda_s^2+\alpha\mathbb{E}\lambda_s\big]\po s\po u\\
  &+6\int_t^{t+\Delta}\int_t^u\int_t^s e^{-(\beta-\alpha)(u+s-2r)}\big[\mathbb{E}\lambda_r^3+2\alpha\mathbb{E}\lambda_r^2+\alpha^2\mathbb{E}\lambda_r\big]\po r\po s\po u
\\
    & + 6\alpha\int_t^{t+\Delta}\int_t^u\int_t^s e^{-(\beta-\alpha)(u-r)}\big[\beta\lambda_\infty\int_t^r\mathbb{E}\lambda_w\po w+\mathbb{E}\lambda_r^2+\alpha\mathbb{E}\lambda_r\big]\po r\po s\po u\\
    & +6(\alpha^2+2\beta\lambda_\infty)\int_t^{t+\Delta}\int_t^u\int_t^s\int_t^r e^{-(\beta-\alpha)(u+s-r-w)}\big[\beta\lambda_\infty\int_t^w\mathbb{E}\lambda_x\po x\big]\po w\po r\po s\po u\\
    & +6(\alpha^2+2\beta\lambda_\infty)\int_t^{t+\Delta}\int_t^u\int_t^s\int_t^r e^{-(\beta-\alpha)(u+s-r-w)}\big[\mathbb{E}\lambda_w^2+\alpha\mathbb{E}\lambda_w\big]\po w\po r\po s\po u\\
    & +6\beta\lambda_\infty\int_t^{t+\Delta}\int_t^u\int_t^s\int_t^r e^{-(\beta-\alpha)(u-s+r-w)}\big[\beta\lambda_\infty\int_t^w\mathbb{E}\lambda_x\po x\big]\po w\po r\po s\po u\\
    & +6\beta\lambda_\infty\int_t^{t+\Delta}\int_t^u\int_t^s\int_t^r e^{-(\beta-\alpha)(u-s+r-w)}\big[\mathbb{E}\lambda_w^2+\alpha\mathbb{E}\lambda_w\big]\po w\po r\po s\po u.
\end{split}
\end{equation*}
In the next step, as before, we take the limit with $t\rightarrow\infty$
and then analytically determine the values of all integrals to
derive \eqref{secondtheorem}.
\QEDA
\vspace{0.8cm}

\subsection{Algorithm to generate the Hawkes process trajectory}
\newpage
\begin{lstlisting}[label={kod},caption={Algorithm to generate the Hawkes process trajectory.}]
simulationNt <- function(alfa, beta, lambda.inf, lambda0, K){
  Ts <- c(0)
  lambdas <- c(lambda0)
  Ns <- c(0)
  for(k in 0:(K-1)){
    D <- 1 + beta*log(runif(1))/(lambdas[k+1]-lambda.inf)
    S2 <- -1/lambda.inf*log(runif(1))
    if(D > 0){
      S1 <- -1/beta*log(D)
      Sk <- min(S1,S2)
    }
    else{Sk <- S2}
    Ts[k+2] <- Ts[k+1] + Sk
    lambdas[k+2] <- (lambdas[k+1]-lambda.inf)*exp(-beta*(Ts[k+2]-Ts[k+1]))+lambda.inf + alfa
    Ns[k+2] <- Ns[k+1]+1
  }
  return(data.frame(Ts, Ns, lambdas))
}
\end{lstlisting}


\end{document}